# Tunable van Hove singularity without structural instability in Kagome metal CsTi$_3$Bi$_5$

Bo Liu[1,#], Minquan Kuang[2,#], Yang Luo[1,#], Yongkai Li[3,4,#], Linwei Huai[1], Shuting Peng[1], Zhiyuan Wei[1], Jianchang Shen[1], Bingqian Wang[1], Yu Miao[1], Xiupeng Sun[1], Zhipeng Ou[1], Yugui Yao[3,4], Zhiwei Wang[3,4,5,*] and Junfeng He[1,*]

[1]Department of Physics and CAS Key Laboratory of Strongly-coupled Quantum Matter Physics, University of Science and Technology of China, Hefei, Anhui 230026, China

[2]Chongqing Key Laboratory of Micro & Nano Structure Optoelectronics, and School of Physical Science and Technology, Southwest University, Chongqing 400715, China

[3]Centre for Quantum Physics, Key Laboratory of Advanced Optoelectronic Quantum Architecture and Measurement (MOE), School of Physics, Beijing Institute of Technology, Beijing 100081, China

[4]Beijing Key Lab of Nanophotonics and Ultrafine Optoelectronic Systems, Beijing Institute of Technology, Beijing 100081, China

[5]Material Science Center, Yangtze Delta Region Academy of Beijing Institute of Technology, Jiaxing, 314011, China

[#]These authors contributed equally to this work.

*To whom correspondence should be addressed:

J.H. (jfhe@ustc.edu.cn), Z.W.(zhiweiwang@bit.edu.cn)

In Kagome metal CsV$_3$Sb$_5$, multiple intertwined orders are accompanied by both electronic and structural instabilities. These exotic orders have attracted much recent attention, but their origins remain elusive. The newly discovered CsTi$_3$Bi$_5$ is a Ti-based Kagome metal to parallel CsV$_3$Sb$_5$. Here, we report angle-resolved photoemission experiments and first-principles calculations on pristine and Cs-doped CsTi$_3$Bi$_5$ samples. Our results reveal that the van Hove singularity (vHS) in CsTi$_3$Bi$_5$ can be tuned in a large energy range without structural instability, different from that in CsV$_3$Sb$_5$. As such, CsTi$_3$Bi$_5$ provides a complementary platform to disentangle and investigate the electronic instability with a tunable vHS in Kagome metals.

Kagome metals AV$_3$Sb$_5$ (A=K, Rb, Cs) have attracted much interest due to the coexistence of multiple exotic orders and states, ranging from superconductivity [1-3], charge density wave (CDW) [4-12], pair density wave [12], stripe order [4], nematic order [13,14], topologically nontrivial states [1,15] and time-reversal symmetry breaking states [11,16-18]. Despite the richness of these phenomena, their underlying mechanisms are still under debate. In principle, either electronic or structural instabilities of a material can drive the system into an ordered state with a lower energy. In AV$_3$Sb$_5$ (A=K, Rb, Cs), electronic instabilities are naturally provided by vHSs in the electron dispersion [19-22], and the structural instabilities are evidenced by the imaginary frequency in the phonon dispersion [23]. As a result, the explanations of the experimentally identified orders are often controversial. For example, the CDW order in AV$_3$Sb$_5$ (A=K, Rb, Cs) has been attributed to either electronic nesting between van Hove singularities [22-25] or electron-phonon coupling [26-30]; the rotational symmetry breaking has been associated with either electronic nematicity [13] or lattice modulation [14,31]. The coexisted instabilities in both electron and lattice degrees of freedom make it very challenging to identify the primary driving mechanism for the various orders in AV$_3$Sb$_5$ (A=K, Rb, Cs). In this regard, the importance of comparative studies in a parallel material system is clear. Theoretical calculations have predicted dozens of materials, which are similar to CsV$_3$Sb$_5$ [32,33]. However, ATi$_3$Bi$_5$ (A=Cs, Rb) is the only material family that has been successfully synthesized recently [34,35].

In this paper, we investigate pristine and Cs surface doped CsTi$_3$Bi$_5$ samples by angle-resolved photoemission spectroscopy (ARPES) and first-principles calculations. The band structure of CsTi$_3$Bi$_5$ is clearly revealed, which shows a clear resemblance to the calculated results. The vHS is well above Fermi level (E$_F$) in the pristine CsTi$_3$Bi$_5$. Surprisingly, the position of the vHS can be easily tuned in a large energy range by Cs surface doping. This property is distinct from that in CsV$_3$Sb$_5$, where the Cs surface doping primarily changes the Sb orbitals but has little effect on the vHS formed by V orbitals. First-principles calculations further reveal the absence of structural instability in both pristine and electron doped CsTi$_3$Bi$_5$. As such, our results establish CsTi$_3$Bi$_5$ as a complementary material platform to CsV$_3$Sb$_5$, in which the electronic instability can be systematically examined without the interference from lattice degree of freedom.

Single crystals of CsTi$_3$Bi$_5$ were grown by a self-flux method with binary Cs-Bi as flux. The raw materials were loaded in an alumina crucible and sealed in an evacuated quartz tube. The tube was heated slowly to 1000 °C and held for 12 h. It was then cooled down to 850°C at a rate of 10 °C /h and to 500°C at a rate of 3°C /h, at which the flux was removed by a centrifuge. The ARPES measurements were carried out at our lab-based ARPES system using 21.2eV photons with a total energy resolution of ~5meV and a base pressure of better than $5 \times 10^{-11}$ torr. The Fermi level was determined by measuring a polycrystalline Au piece in electrical contact with the samples. First-principles calculations were performed by using the VASP software package. The details of the calculations and the related parameters are described in the supplemental material.

The crystal structure of $CsTi_3Bi_5$ is similar to that of $CsV_3Sb_5$ (Fig. 1a). The Ti sublattice forms a Kagome net, which is interwoven with a hexagonal net of Bi atoms in the same plane. The measured band structure of $CsTi_3Bi_5$ is shown in Fig. 1b, which bears a clear resemblance to that of the first-principles calculations (Fig. 1d). Due to the layered nature of the material, a projected in-plane Brillouin zone (BZ) is used for the description. The electronic structure near the $\bar{\Gamma}$ point is dominated by an electron-like band (labelled as $\alpha$ band, hereafter), giving rise to a circular Fermi surface sheet (Fig. 1c). Multiple hole-like bands are observed around the $\bar{M}$ point (labelled as $\beta$, $\gamma$ and $\delta$ band, respectively). They are associated with the hexagonal, flower-like and diamond-like Fermi surface sheet, respectively (Fig. 1c). A Dirac-like crossing can be seen at $\bar{K}$, and a triangular Fermi surface sheet is observed around the $\bar{K}$ point (Fig. 1c). The characteristic vHS of the Kagome lattice is shown at $\bar{M}$ point in the calculation (Fig. 1d). Nevertheless, it locates at ~150 meV above $E_F$, which cannot be probed by the photoemission measurements (Fig. 1d).

After revealing the overall electronic structure of the $CsTi_3Bi_5$, we now investigate the doping evolution via *in situ* surface deposition of Cs atoms. As shown in Fig. 2a, the electron-like band ($\alpha$ band) around $\bar{\Gamma}$ shows a moderate change as a function of Cs doping (Fig. 2a). The distance between the two Fermi momenta ($k_{F1}$ and $k_{F2}$) increases slightly with doping (Fig. 2b). On the contrary, the energy bands around $\bar{M}$ exhibit more significant changes as a function of Cs doping (Fig. 2d). In particular, the $\gamma$ and $\delta$ bands present a clear downward shift, echoing the expected electron doping with Cs surface deposition. We note that the top of these hole-like bands starts to appear with sufficient Cs doping [Fig. 2d(v)], indicating that the vHS is in the vicinity of $E_F$. This is also evidenced by the enhanced electron density of states at $E_F$ in this momentum region (Fig. 2e, f). The integrated energy distribution curve (EDC) around the $\bar{M}$ point shows a negligible peak in the pristine $CsTi_3Bi_5$, as the vHS is well above $E_F$ (Fig. 2e). However, the peak intensity increases significantly with Cs doping, demonstrating the boost of low energy electron density of states as the vHS approaches $E_F$ (Fig. 2f). In order to quantitatively unveil the vHS, the Cs surface doping is reproduced at an elevated temperature (T=200K), where the thermal population of electrons enables a complete examination of the fine features around $E_F$. As shown in Fig. 3a-d, the vHS is indeed shifted downward with doping. On the sufficiently doped sample, the flat dispersion of the vHS can be clearly identified in the vicinity of $E_F$ (Fig. 3b, d). These results are quantitatively extracted from the data and summarized in Fig. 3f. Orbital-resolved calculations have also been carried out, which illustrate that the electron-like α band around $\bar{\Gamma}$ is dominated by Bi $P_z$ orbital, whereas the vHS is primarily associated with Ti $d_{x^2-y^2}$ orbital (Fig. 3e). These observations have collectively depicted an integrated picture of orbital selective movements of the energy bands with Cs doping -- the vHS with Ti $d$ orbitals can be tuned in a large energy range, whereas the electron-like α band with Bi $P$ orbitals remains less sensitive to the doping process. This is distinct from the evolution in $CsV_3Sb_5$, where the vHS with V $d$ orbitals shows little change with Cs surface doping, but the electron-like band with Sb $P$ orbitals shifts ~240 meV in energy [36].

Next, we examine the lattice degree of freedom in the $CsTi_3Bi_5$ crystal. We have followed the idea in $CsV_3Sb_5$ [23], and calculated the change of total energy in $CsTi_3Bi_5$, assuming that the lattice is breathing in and out towards the potential Star of David (SD) and inverse Star of David (ISD) structures (Fig. 4a). In $CsV_3Sb_5$, either SD or ISD structure shows a lower total energy than that of the Kagome structure (Fig. 4b), leading to structural instabilities of the material [23]. On the contrary, the Kagome structure in $CsTi_3Bi_5$ exhibits the lowest total energy, demonstrating the absence of structural instability (Fig. 4c). This result remains solid when electron doping is considered in the $CsTi_3Bi_5$ system (Fig. 4d). Phonon spectra are also calculated for both pristine and electron doped $CsTi_3Bi_5$ (Fig. 4e). The absence of imaginary frequency echoes a stable Kagome structure in $CsTi_3Bi_5$.

Finally, we discuss the implications of our observations. The tunable vHS and the absence of structural instabilities make $CsTi_3Bi_5$ a complementary material platform to compare with $CsV_3Sb_5$. Without the interference from lattice, one can systematically examine the electronic instabilities associated with the vHS. For example, the CDW order is absent in the pristine $CsTi_3Bi_5$ [34,35], and no CDW gap is observed in our ARPES measurements (see supplemental material). When the vHS is tuned to the vicinity of $E_F$, the photoemission spectra remain gapless at low temperature (see supplemental material). There results indicate that the electronic nesting between vHSs at $\overline{\text{M}}$ points is insufficient to drive a CDW order in the Kagome metal. The structural instabilities in CsV$_3$Sb$_5$ play an essential role in this context. On the other hand, nematic order might be driven by pure electronic interactions, as it has been reported in both CsV$_3$Sb$_5$ and CsTi$_3$Bi$_5$ [13,14,37,38]. It would be interesting to further explore how the tunable vHS in CsTi$_3$Bi$_5$ would interact with the nematic order and other potential electronic orders in the system.

In summary, we have revealed the electronic structure of pristine and Cs surface doped CsTi$_3$Bi$_5$ samples. The Cs deposition induces an overall electron doping to the material, but the energy bands exhibit an orbital dependent movement with doping. Among them, the vHS can be tuned in a large energy range. First-principles calculations demonstrate that the Kagome structure remains stable in both pristine and electron doped CsTi$_3$Bi$_5$. These results establish a unique path to disentangle the electronic instability from that of the lattice, and to examine its relationship with the various exotic phenomena in Kagome metals.

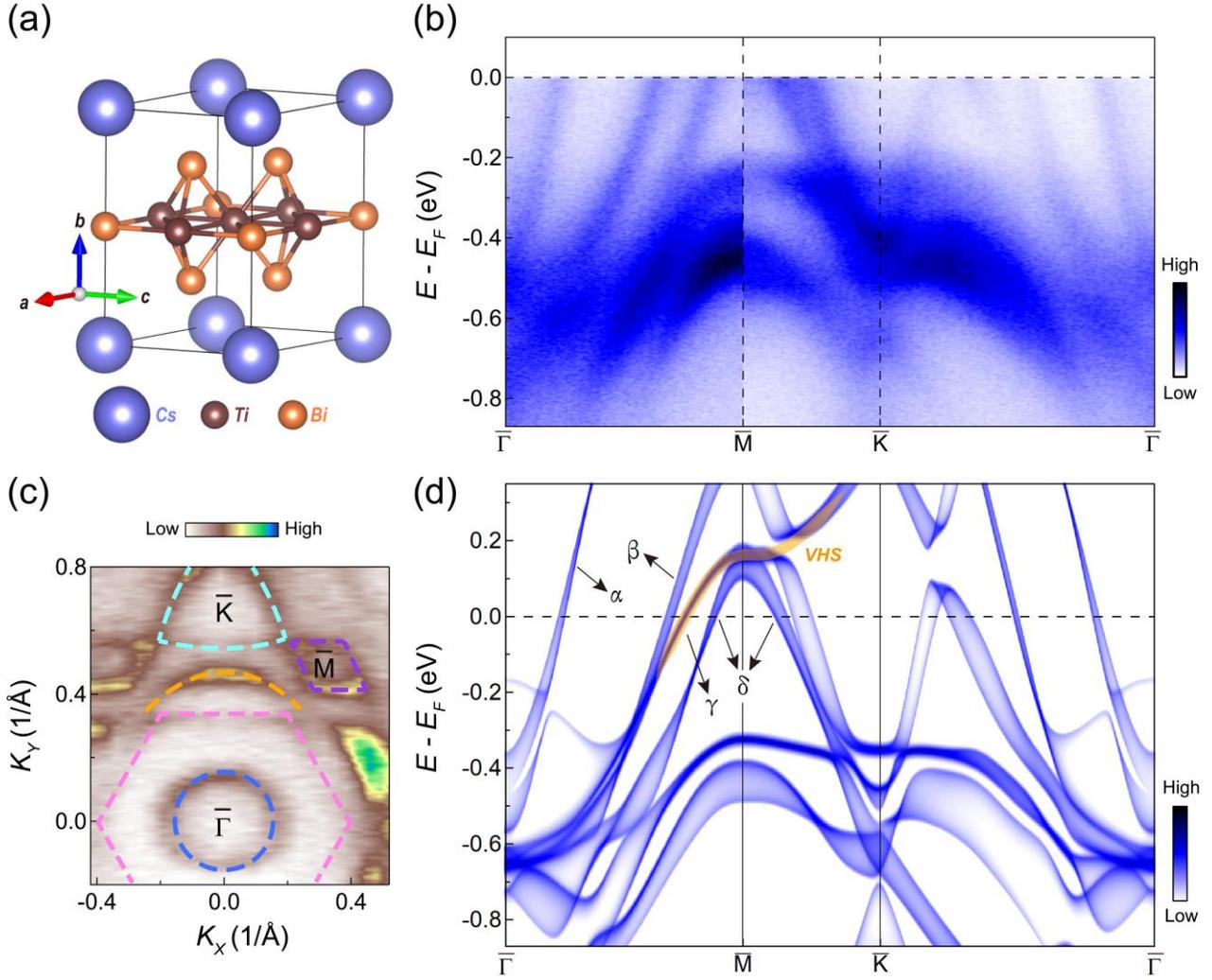

**FIG. 1. Electronic structure of CsTi$_3$Bi$_5$.** (a) Crystal structure of CsTi$_3$Bi$_5$. (b-c) Photoelectron intensity plot along $\overline{\Gamma}$-$\overline{M}$-$\overline{K}$-$\overline{\Gamma}$ (b) and Fermi surface (c) of CsTi$_3$Bi$_5$ measured with 21.2 eV photons at 7K. The dashed lines in (c) are a guide to the eye. (d) The bulk band structure of CsTi$_3$Bi$_5$ obtained from first-principles calculations with spin-orbital coupling included. The electron-like band around $\overline{\Gamma}$ and the hole-like bands near $\overline{M}$ are labelled as α, β, γ and δ band, respectively. The orange shade is an eye-guide for the vHS.

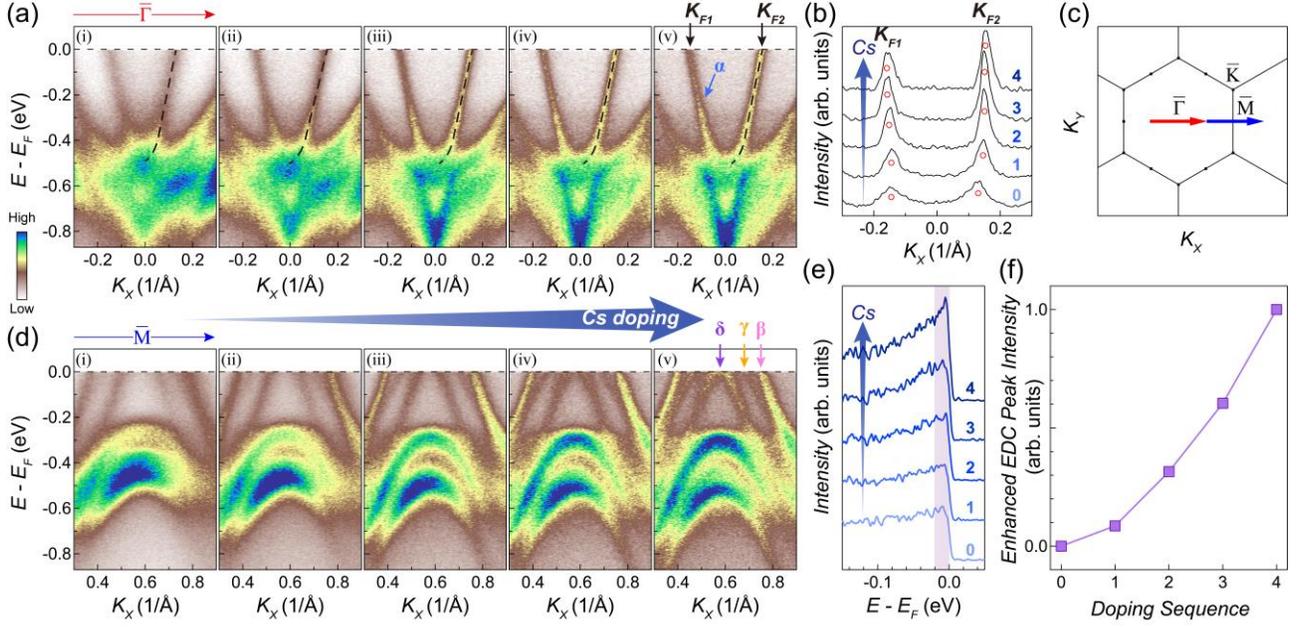

**FIG. 2. Evolution of the electronic structure with Cs surface doping at 7K.** (a) Photoelectron intensity plot of the band structure around $\bar{\Gamma}$ as a function of continuous Cs doping on the same sample. The results at doping sequences 0-4 are shown in (i-v), respectively. Doping sequence 0 indicates the pristine $CsTi_3Bi_5$ sample. The dashed lines are a guide to the eye. (b) Momentum distribution curves (MDCs) at $E_F$ extracted from (a). (c) The projected in-plane BZ and the momentum locations of the cuts. (d) Same as (a), but for the band structure around $\bar{M}$. (e) Integrated EDC around the $\bar{M}$ point in (d). The numbers 0-4 denote the doping sequences. (f) Enhanced EDC peak intensity around $\bar{M}$ as a function of the doping sequence. The absolute EDC peak intensity at the doping sequence x (x=0-4) is calculated by integrating the area between -20meV and $E_F$ of the corresponding EDC, and labelled as $I_x$. The enhanced EDC peak intensity is defined as $(I_x-I_0)/(I_4-I_0)$.

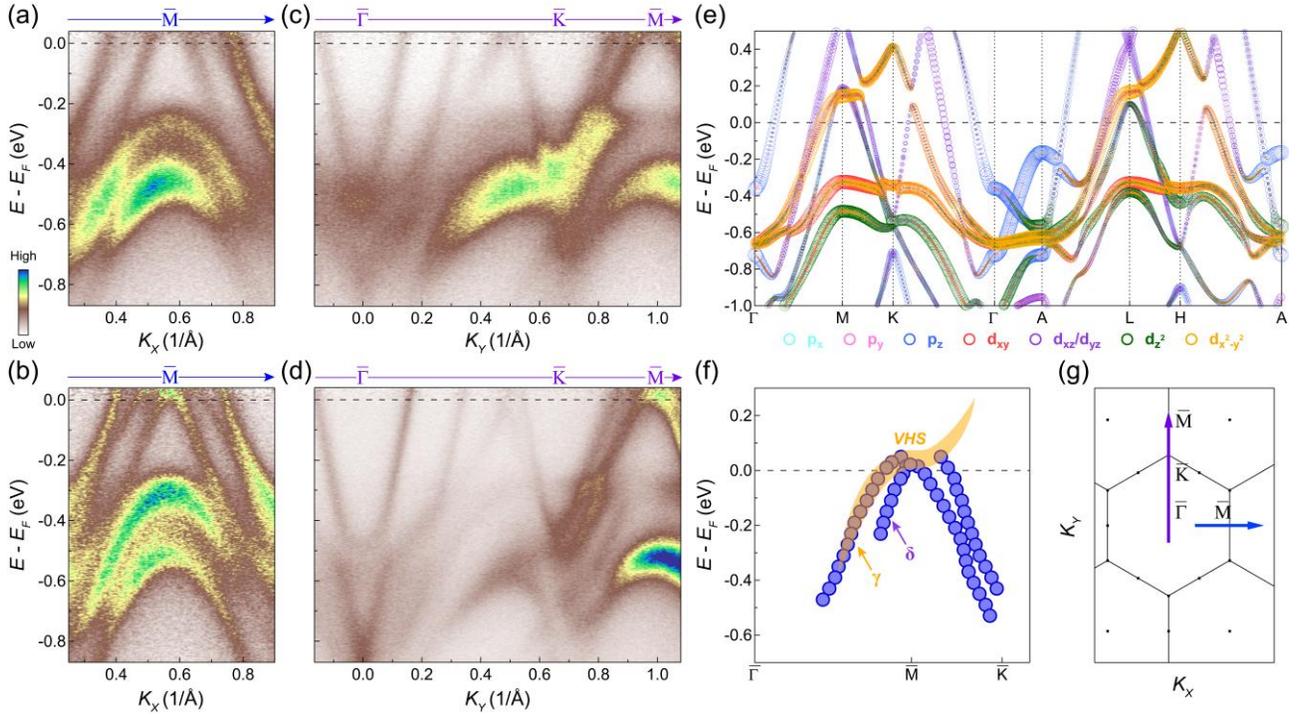

**FIG. 3. Doping evolution of the vHS.** (a-b) Photoelectron intensity plot of the band structure around $\overline{M}$ (along the $\overline{\Gamma}$-$\overline{M}$-$\overline{\Gamma}$ direction) before (a) and after (b) the Cs surface doping, measured at 200K. (c-d) Same as (a-b), but measured along the $\overline{\Gamma}$-$\overline{K}$-$\overline{M}$ direction. (e) Orbital-resolved band structure obtained by first-principles calculations. (f) Quantified dispersion of the γ and δ bands near $\overline{M}$ after sufficient Cs doping, extracted from (b) and (d). The orange shade is an eye-guide for the vHS. (g) Momentum locations of the cuts in the BZ.

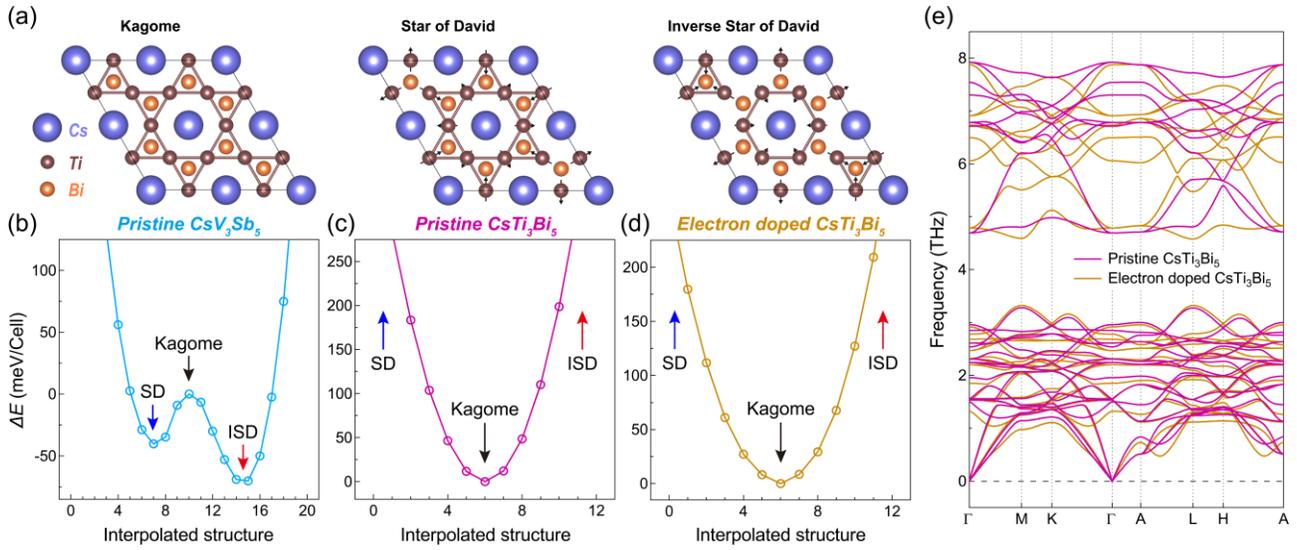

**FIG. 4. Calculated total energy profiles and phonon spectra.** (a) The 2×2 supercells for Kagome structure, Star of David structure and Inverse Star of David structure. The black arrows indicate the lattice distortion due to the breathing mode. (b-d) Total energy as a function of the interpolated structure in pristine $CsV_3Sb_5$ (b), pristine $CsTi_3Bi_5$ (c), and electron doped $CsTi_3Bi_5$ (d). (e) Calculated phonon spectra along the high-symmetry directions in pristine $CsTi_3Bi_5$ (magenta line) and electron doped $CsTi_3Bi_5$ (orange line).